\documentstyle{Europhys}

\def\And{{\rm and\ }}

\def\drm{{\rm d}}

\def\stars{\bigskip\centerline{***}\medskip}

\newif\ifboo \boofalse

\def\Review#1{\boofalse{\it #1},}
\def\Name#1{{\sc #1},}
\def\Vol#1{\ifboo Vol. {\bf #1}\else{\bf #1}\fi}
\def\Year#1{\ifboo #1\else(#1)\fi}

\def\Page#1{\ifboo {\rm p. #1}\else{\rm #1}\fi}
\input epsf.def
\def\Int{\rm{Int }}

\def\eq{\begin{equation}}
\def\en{\end{equation}}
\def\eqy{\begin{eqnarray}}
\def\eny{\end{eqnarray}}

\begin{document}
\euro{}{}{}{}
\Date{}
\shorttitle{T. C. HALSEY: THE BRANCHING STRUCTURE ETC.}
\title{\bf The branching structure of diffusion-limited aggregates}
\author{T. C. Halsey\inst{1}}
\institute{\inst{1}Exxon Research and Engineering, Route 22 East,
Annandale, N.J. 08801, U.S.A. \\}
\rec{}{}
\pacs{\Pacs{61}{43.Hv}{Fractals; macroscopic aggregates (including
diffusion-limited aggregates)}
\Pacs{47}{53.+n}{Fractals}
\Pacs{05}{40.+j}{Fluctuation phenomena, random processes, and Brownian
motion}}
\maketitle
\begin{abstract} 
I analyze the topological structures generated by diffusion-limited
aggregation (DLA), using the recently developed ``branched growth model".
The computed bifurcation number $B$ for DLA in two dimensions is $B \approx
4.9$, in good agreement with the numerically obtained result of $B \approx
5.2$. In high dimensions, $B \to 3.12$; the bifurcation ratio is thus a
decreasing function of dimensionality. This
analysis also determines the scaling properties of the ramification matrix,
which describes the hierarchy of branches.
\end{abstract}

Nature creates an astonishing variety of branched structures. Some of
these structures, such as trees, are created by biological systems;
others, such as river networks, are created principally by physical
phenomena. The simplest and best-understood physical model for the
formation of branched structures is diffusion-limited aggregation (DLA),
introduced over fifteen years ago \cite{dla}. The complex fractal
structures generated by this model are seen in natural systems whose growth
is controlled by diffusive processes \cite{expt}.  

Physicists have concentrated on understanding and
predicting the fractal and
multifractal properties of
DLA \cite{pietronero,prl}.
However, several authors have advanced an alternative approach,
focussing on the topological self-similarity of DLA clusters, as measured by
quantities typically used to describe river networks and other branched
structures
\cite{norway,vanni}. This Letter is devoted to the
computation of these quantities, using the ``branched growth model" which
this author and his collaborators have exploited to compute a number of
properties of DLA \cite{pra}. Since this model is based on an analysis of
the competition of branches in a hierarchical structure, it is peculiarly
suited to the computation of topological quantities related to that
structure.

The growth rule for DLA can be defined inductively: introduce
a random walker at a large distance from an $n$ particle cluster, which
sticks irreversibly at its point of first contact with the cluster,
thereby forming the $n+1$ particle cluster. Clusters grown in this way
have an intricate branched structure, in which prominent branches screen
internal regions of the cluster, preventing them from growing further. In
addition, they contain no loops, since the particle closing a loop would
have to attach to two pre-existing particles. The scaling of the radius of
the cluster $r$ with the number of particles
$n$ determines the fractal dimension $D$ of the cluster, $n \propto r^D$.

\begin{figure}

\centerline{
\epsfbox{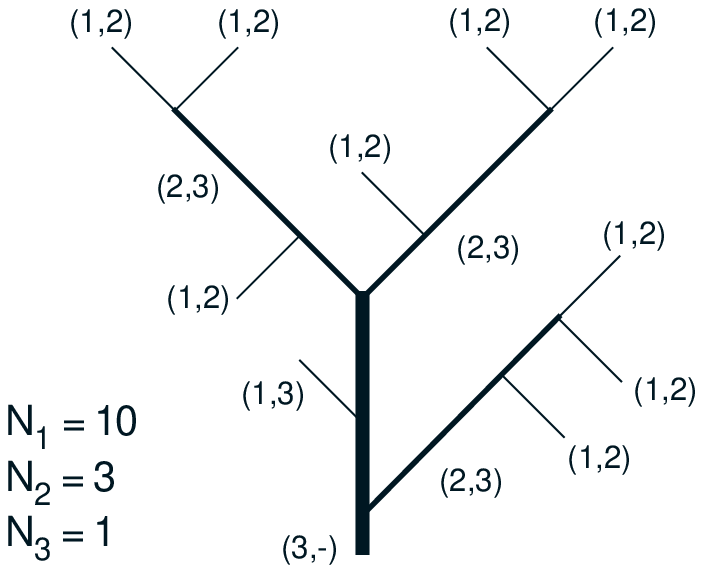}
}

\caption{A rooted tree, showing the Strahler indices and parent Strahler
indices of each branch. The first number is the Strahler index of a branch,
and the second is the Strahler index of the branch it meets when it loses
its identity. The total number of index 1 branches is $N_1 = 10$, of index
2 branches is $N_2 = 3$, and of index 3 branches is $N_3 = 1$.}
\label{1}
\end{figure}

A number of quantities have been developed to describe the topological
properties of river networks, which like DLA clusters have no loops. The
most fundamental such quantity was introduced by Strahler, as a refinement
of a scheme proposed by Horton
\cite{strahler}. Consider a rooted tree structure. The leaves of
the structure are assigned the Strahler index $i=1$. The index $i$ of any
other branch, which bifurcates into two branches with Strahler indices
$i_1$ and
$i_2$, is determined by 
\eq
i = \cases{ i_1 + 1, & if $i_1 = i_2$ \cr
\max(i_1,i_2), & if $i_1 \ne i_2$. \cr}
\label{eq:strahler}
\en
Thus a change in the index of a branch occurs only when it joins another
branch of roughly the same size (see fig.\ \ref{1}). 

Once we have defined the Strahler index of each branch, we can also obtain
$N_i$, the number of branches of index $i$. Note that a branch
with several side-branches of lower index counts as a single branch for
this numeration (fig.\ \ref{1}). The bifurcation ratios $B_i$ are then
defined by $B_i = N_i / N_{i+1}$; by definition $B_i \ge 2$. River networks
are often topologically self-similar, with $B_i \approx B$ and $3 < B <5$,
even though they do not display conventional geometrical fractality, since
they extend close to every point in a basin. DLA clusters in two dimensions
also exhibit topological self-similarity, with $B_i \approx 5.2$ for $i > 1$
\cite{norway,vanni,ossadnik}.

If the characteristic
size (defined in whatever way is convenient) of branches of index $i$ is
$L_i$, then we can define a length ratio $A_i$ by $A_i = L_i / L_{i+1}$.
Self-similarity would then demand that $A_i \approx A$, with the fractal
dimension given by $D = \ln B / \ln A$ \cite{norway}.

The Tokunaga-Viennot ordering refines this Strahler ordering, by defining
two indices $i,j$ for every branch \cite{tokunaga,viennot}. The first index
$i$ is the Strahler index of the branch, while the second index $j$ is the
Strahler index of its ``parent" (see fig.\ \ref{1}). By definition, $j >
i$. The ramification matrix
$R_{ij}$ is the number of branches with the index pair
$ij$; clearly $\sum_j R_{ij} = N_i$. By contrast,
$\sum_i R_{ij}$ gives the total number of side-branches
attached to index
$j$ branches.  Since sidebranches are separated by microscopic distances
on any particular branch, 
\eq
\sum_i R_{ij}/N_j \propto L_j .
\label{eq:A}
\en

In addition to river networks,
these quantities have also been used in computer graphics, and in the
mathematical analysis of random binary trees \cite{viennot,penaud}. These
latter have
$B=4$ asymptotically --we shall see below that this result is
bracketed by the DLA results for $d=2$ and $d \to \infty$.

\section{The branched growth model}
The branched growth model for DLA is based upon a quantitative analysis
of the competition between neighboring branches (indexed by 1, 2) that
join at a particular node or branch point. Let the number of particles of
branches 1 and 2 be
$n_1$, $n_2$, and their total growth probability be $p_1$, $p_2$.
Let us define normalized parameters $x$, $y$ by $x=p_1/(p_1+p_2)$ and
$y=n_1/(n_1+n_2)$. The total number of particles in the two branches is
$n_b \equiv n_1 + n_2$. We can write a simple equation of motion for
$y$:
\eq
\frac{\drm y}{\drm \ln n_b} =   \frac{\drm
n_1}{\drm n_b} - \frac{n_1} {n_b} = x - y .
\label{eq:ydot}
\en
Note that the right-hand side is a function of $x$ and $y$, but not
explicitly of
$n_b$. I have computed a similar equation of motion for $x$,
\eq
\frac{\drm x}{\drm \ln n_b} = G(x,y) ,
\label{eq:xdot}
\en
by a renormalization method for two-dimensional DLA:
$G(x,y)$ is a function of $x$ and $y$ alone, but is too complicated to give
here
\cite{prl}. 

Solving these differential equations for $x,y$, one
generates a competition dynamics for these two branches in the
space of their relative growth probabilities and masses, $(x,y)$.  This
competition dynamics is symmetrical about a hyperbolic unstable fixed point
at $(x,y)=(\frac{1}{2},\frac{1}{2})$; as the number of particles
$n_b$ in the two branches increases, $(x,y)$ flows to a stable
fixed point at either $(0,0)$ or $(1,1)$. The eigenvalue of the
unstable direction at the central, unstable fixed point is termed $\nu$,
so that for trajectories starting near this fixed point
\eq
y(n_b) - \frac{1}{2} \propto n_b^{\nu} ,
\label{eq:unstable}
\en
and similarly for $x$. The hyperbolic nature of the  unstable fixed point
implies that there is a stable manifold flowing into it, and an unstable
manifold flowing out of it and into the stable fixed points.

In order to understand the statistical distribution in
$x-y$ space of branch pairs, we must understand not only their evolution but
also their birth.  I assume that branch pairs are created by microscopic
tip-splitting events, which are indifferent to the large scale structure
created by eqs.\ (\ref{eq:ydot},\ref{eq:xdot}). Thus those
branch pairs born in the neighborhood of the unstable fixed point will
have a roughly constant initial distribution. These branch pairs will
remain ``active", i.e., far from the stable fixed points, until $n_b \gg
1$, and thus they will determine the large-scale structure of the cluster.
Let us suppose that the distance from the unstable fixed point (or more
properly, the stable manifold) at birth, where $n_b
\sim 1$, is given by
$\epsilon^{\nu}$, thus defining $\epsilon$. 
A probability distribution for $\epsilon$ of $\rho(\epsilon) d
\epsilon =
\rho_0
\epsilon^{\nu-1} d \epsilon$ for $\epsilon \ll 1$ (with $\rho_0$ a
constant) implies the desired constant distribution near the stable manifold
in the $x-y$ plane. A simple argument then gives
$\nu=D^{-1}$ \cite{pra}.  

Introducing a new parameter
$\eta=\epsilon n_b$, the branch competition determines a
parameterization
$x(\eta)$, $y(\eta)$ of the unstable manifold; these functions obey eqs.\
(\ref{eq:ydot},\ref{eq:xdot}) with $x(0) = \frac{1}{2}$ and $y(0) =
\frac{1}{2}$. For simplicity we choose
$x,y$ so that $\lim_{\eta \to \infty} x(\eta),y (\eta) = 0$; thus for
large $n_b$
$(x,y)$ refers to the weak member of the branch pair, and $(1-x,1-y)$ to
the strong member.

We know the function $G(x,y)$ in two dimensions, but not in higher
dimensions. In this case, a useful approximation is given by ``model Z",
for which the unstable manifold of the equations of motion is given by a
straight line which commences at the unstable fixed point and continues
until it reaches
$x=0$ (or $x=1$) \cite{pra}. It then moves parallel to the $y$-axis into the
stable fixed point at either $(x,y) = (0,0)$ or $(x,y)=(1,1)$. Thus model Z
is given by $x(\eta) = \frac{1}{2} - \eta^{\nu}$ for $\eta^{\nu} <
\eta_c^{\nu}
\equiv \frac{1}{2}$, and eq.\ (\ref{eq:ydot}) implies
\eq
y(\eta)  = \cases{{1/2} - (\eta^{\nu}/1 + \nu) & $\eta < 
\eta_c$
\cr {\bar \eta}/{\eta} &
$\eta > \eta_c$ \cr}  
\label{eq:zy}
\en
where $\bar \eta= (1/2)^{1 + \nu^{-1}}
\nu/(1+\nu)$. Model Z specifies a one-parameter family of unstable
manifolds; it is possible to determine which of these best approximates DLA
as a function of spatial dimensionality by a self-consistency argument
\cite{pra}. 

\section{Bifurcation numbers}
Consider a branch with a constant bifurcation number of $B$ whose root has a
Strahler index of $I$. We term the Strahler {\it number} of this branch to
be $I$, the Strahler index of its root. For a DLA cluster, the total number
of particles in the cluster $n$
will be proportional to the total number of elementary sub-branches $n_e$,
with
$n_e \approx B^{I-1}$. Inverting this relationship,
we find
\eq
I(n) - 1 \approx \ln n / \ln B \equiv \alpha \ln n ,
\label{eq:I}
\en
where the relation is approximate because the Strahler number is by
definition an integer. Note that we are assuming that the Strahler number
of a branch is determined uniquely by the number of particles in the
branch. In practice, the Strahler number of branches of given size will
fluctuate in the ensemble of DLA clusters; little is known about such
fluctuations.

The main branch of the cluster is defined by starting at
the root and at every branching choosing the stronger of the two
sub-branches. This main branch eventually ends, as do all branches, in an
elementary sub-branch tip. Equation (\ref{eq:I}) implies that
\eq
\alpha \ln n = \sum_k p_k ,
\label{eq:fund}
\en
where the sum is over all nodes of the main branch, and $p_k$ is the
probability that the Strahler index of the main branch increases by 1 at
the $k$'th node, as will be the case if the two sub-branches at this node
are close enough in size to have the same Strahler number.

Suppose that the Strahler
number of a branch is taken to be exactly 
\eq
I(n) = \Int( \alpha \ln n) ,
\en
where $\Int$ indicates the integer part. Now consider two sub-branches of
the main branch of size
$n_1$ or
$n_2$. If
$\vert
\ln n_2 -\ln n_1 \vert>
\alpha^{-1}$, then they have different Strahler numbers. Otherwise, the
probability $p(n_1,n_2)$ that they have the same Strahler number is
\eq
p(n_1,n_2) = 1 -\alpha \vert \ln n_2 -\ln n_1 \vert .
\en
Thus we can write
\eq
p_k = \left \langle \int_0^{\tilde \epsilon} d \epsilon_k \; \rho
(\epsilon_k)
\bigg
( 1 -
\alpha  \big \{ \ln [1-y(\epsilon_k n_k)] -\ln y(\epsilon_k n_k) \big \}
\bigg
)
\right
\rangle ,
\label{eq:logy}
\en
where $n_k \gg 1$ is the total number of particles below (away from the
root) the $k$'th node, and $\tilde \epsilon$ is defined by $\ln
[1-y(\tilde
\epsilon n_k)] -\ln y(\tilde \epsilon n_k) = \alpha^{-1}$. The brackets
$\langle \cdot \rangle$ indicate averaging over the random variables at all
nodes of the main branch excluding node $k$, whose average is indicated
explicitly. For $n_k \gg 1$, we can use the small
$\epsilon$ result $\rho(\epsilon) = \rho_0 \epsilon^{\nu-1}$ to obtain
\eq
p_k = \left \langle \frac{\rho_0}{n_k^{\nu}} \right \rangle\int_0^{\tilde
\eta} d \eta \; \eta^{\nu-1} \bigg ( 1 -
{\alpha} \big \{ \ln [1-y(\eta)] -\ln y(\eta) \big \} 
 \bigg ) \equiv \left \langle \frac{\rho_0}{n_k^{\nu}} \right
\rangle \Gamma(\alpha) ,
\label{eq:pk}
\en
with $\ln [1-y(\tilde \eta)] -\ln y(\tilde \eta) = \alpha^{-1}$.
In
ref.\ \cite{hhd} we obtained the useful identity
\eq
\left \langle \sum_k \frac{\rho_0}{n_k^{\nu}} \right \rangle = -\left \{
\int_0^{\infty} d
\eta \;
\eta^{\nu-1} \ln [1 - y (\eta) ]  \right\}^{-1} \ln n \equiv
\lambda
\ln n .
\label{eq:sum}
\en
Thus, combining eqs.\ (\ref{eq:fund},\ref{eq:pk},\ref{eq:sum}), we obtain
\eq
\alpha = \lambda \Gamma(\alpha) ,
\label{eq:alpha}
\en
allowing solution for $\alpha$.

For DLA in two dimensions, for which $y(\eta)$ is known,
solving eq.\ (\ref{eq:alpha}) gives $\alpha \approx 0.628$,
or
$B \approx 4.92$, in reasonable agreement with the numerical result of $B
\approx 5.2$
\cite{norway,vanni}.  In higher dimensionalities, we must rely upon model Z.

It is interesting to consider model Z in the limit $\nu \to 0$, or $D=
\nu^{-1}
\to
\infty$. In this limit, $\int d \eta \; \eta^{\nu-1} \ln [1-y (\eta)]$ is
dominated by $\eta \ll \eta_c$, and
\eq
\lambda = \frac{2 \nu } {1 - \ln
2  + O(\nu^2)} .
\label{eq:lambda}
\en
Computing $\Gamma (\alpha)$ with $y(\eta)$ given by eq.\
(\ref{eq:zy}) yields
\eq
\Gamma \left (\alpha \equiv \frac{1}{\ln B} \right ) = - \frac{1 + \nu
}{2 \nu} \frac{1}{\ln B} \ln
\left [ 1 -
\left ( \frac{B-1}{B+1}\right )^2 \right ] .
\en
and eq.\ (\ref{eq:alpha}) gives
\eq
\lim_{\nu \to 0} B = \frac{1+ \sqrt{1-\exp(\ln 2 - 1)}}{1-
\sqrt{1-\exp(\ln 2 - 1)}} \approx 3.12 .
\label{eq:limit}
\en
Our somewhat surprising result is that $B$ is a {\it
decreasing} function of dimensionality. A simple extension of the
above calculation predicts $B \to \infty$ as dimensionality $d \to 1$.

\section{Ramification matrices}
To compute ramification matrices, we must determine the probability that a
branch of index $i_1$ will have a side-branch of index $i_2$. At a
particular node, say the $k$'th node, we can say approximately that 
\eq
i_2-i_1 = I \mbox{  if  } \frac{2 I - 1}{2 \alpha} < - \ln y_k <
\frac{2 I + 1}{2 \alpha} .
\en
for $I \ge 1$. Let us define $\eta_{I}$ by
\eq
\ln y(\eta_{I}) = - \frac{2 I -1}{2 \alpha} .
\en

A branch with $n$ particles will have a total length
$\sim n^{1 / D} = n^{\nu}$. Sidebranches are separated by microscopic
distances, thus the total number of sidebranches of a branch of size
$n$ is also
$\propto n^{\nu}$. The  total number of branches of order $i$ is
$\propto B^{-i}$, leading to
\eq
R_{ij; j-i=I} \propto  B^{-i} \; n^{\nu} \frac{\rho_0}{n^{\nu}}
\int_{\eta_{I}}^{\eta_{I+1}} d \eta \; \eta^{\nu-1} \propto
B^{-i} \left ( \eta_{I+1}^\nu - \eta_{I}^\nu \right ) .
\en
For $\eta$ sufficiently large, $x(\eta) \approx 0$. To further simplify
matters, we suppose that $x=0$ exactly for $\eta > \eta_I$, as would be
the case in model Z for $\eta_I > \eta_c$.  Then eq.\ (\ref{eq:zy}) gives
$y(\eta) = \bar \eta /\eta$, and
\eq
\eta_{I} = \bar \eta \exp \left ( \frac{2 I - 1 }{ 2 \alpha}
\right ) ,
\en
which we expect to be approximately true for non-model Z trajectories.
Thus
\eq
R_{ij; j-i=I} \propto B^{-i} \exp \left ( \nu I / \alpha \right ) =
B^{\nu I - i} .
\en
This result is valid in arbitrary spatial dimensionality. In two
dimensions, we obtain $B^{\nu} \approx 2.6$, in agreement with
ref. \cite{ossadnik}, which claims $R_{ij; j-i=I} \propto (2.7)^I$; however,
since these latter results are based on a variant of the Strahler ordering,
it is not clear that they are relevant.  Comparing with eq.\ (\ref{eq:A}),
we see that $A=B^{\nu}$ and $\ln B/ \ln A = \nu^{-1} = D$, as predicted
\cite{norway}.

\stars

I am grateful to Iddo Yekutieli for bringing this problem to my attention,
and to Don Turcotte for reminding me of it on several occasions
over two years.  I am also grateful to the Newton Institute, Cambridge,
for their hospitality during a visit which initiated this work.

\end{document}